\documentclass[sigplan,screen]{acmart}

\AtBeginDocument{%
  }

\copyrightyear{2025}
\acmYear{2025}
\setcopyright{rightsretained}
\acmConference[ASPDAC '25]{30th Asia and South Pacific Design Automation Conference}{January 20--23, 2025}{Tokyo, Japan}
\acmBooktitle{30th Asia and South Pacific Design Automation Conference (ASPDAC '25), January 20--23, 2025, Tokyo, Japan}\acmDOI{10.1145/3658617.3697625}
\acmISBN{979-8-4007-0635-6/25/01}

\acmConference[ASPDAC'25]{Asia and South Pacific Design Automation Conference}{January 20--23, 2025}{Tokyo, Japan}

\usepackage{amssymb}

\usepackage{threeparttable} % Include the package
\usepackage{cite}
\usepackage{amsmath,amsfonts}
\usepackage{algorithmic}
\usepackage{graphicx}
\usepackage{textcomp}
\usepackage{caption}
\usepackage{subcaption}
\usepackage{caption}
\usepackage{siunitx}
\usepackage{tcolorbox}
\usepackage{xcolor}
\usepackage{multirow}
\usepackage{siunitx}
\usepackage{boldline} 
\usepackage{subcaption,mwe}
\usepackage{enumitem}
\usepackage{soul}
\usepackage{tabularx}
\usepackage{booktabs,chemformula}
\usepackage{mathtools}
\usepackage{makecell}
\usepackage{adjustbox}
\definecolor{lightblue}{RGB}{173,216,230}
\usepackage{booktabs}
\usepackage{siunitx}
\usepackage{makecell}
\usetikzlibrary{shapes.multipart}
\usepackage{tikz}
\usepackage{soul}
\usetikzlibrary{calc} 
\usetikzlibrary{arrows, decorations.markings,positioning,backgrounds,shapes}
\definecolor{EMP}{HTML}{77DD77} % Green1
\definecolor{NOR}{HTML}{06500C} % Green2

\usepackage{threeparttable} % Include the package
\usepackage{booktabs} % For prettier tables
\usepackage{array}
\usepackage{multirow} % Required for multirow feature
\usepackage{caption} % For better control over table captions
\usepackage{booktabs}
\usepackage{graphicx} % For \resizebox
\usepackage{booktabs} % For professional looking tables
\usepackage{multirow} % For multirow feature in tables
\usepackage{threeparttable} % For footnotes in tables
\usepackage{tabularx}
\usepackage{amsmath}
\usepackage{lipsum}
\usepackage{tikz}
\usepackage{cite}
\usepackage{amsmath,amsfonts}
\usepackage{algorithmic}
\usepackage{graphicx}
\usepackage{textcomp}
\usepackage{caption}
\usepackage{subcaption}
\usepackage{multirow}
\usepackage{boldline} 
\usepackage{subcaption,mwe}
\usepackage{soul}
\usepackage{tabularx}
\usepackage{booktabs,chemformula}
\usepackage{mathtools}
\usepackage{makecell}
\usepackage{adjustbox}
\usepackage{makecell}
\usepackage{threeparttable} % Include the package
\usepackage{booktabs} % For prettier tables
\usepackage{array}
\usepackage{multirow} % Required for multirow feature
\usepackage{caption} % For better control over table captions
\usepackage{booktabs}
\usepackage{graphicx} % For \resizebox
\usepackage{booktabs} % For professional looking tables
\usepackage{multirow} % For multirow feature in tables
\usepackage{threeparttable} % For footnotes in tables
\usepackage{tabularx}
\usepackage{soul}
\usetikzlibrary{calc} 
\usetikzlibrary{arrows, decorations.markings,positioning,backgrounds,shapes}
\usepackage{tcolorbox}
\usepackage{xcolor} % for color definitions

\definecolor{pastelblue}{HTML}{aec6cf}
\definecolor{grayframe}{HTML}{d8d8d8}
\definecolor{pastelgreen}{HTML}{77dd77}
\definecolor{whiteframe}{HTML}{ffffff}
\definecolor{pastelblue}{HTML}{aec6cf}
\definecolor{grayframe}{HTML}{d8d8d8}
\definecolor{pastelgreen}{HTML}{77dd77}
\definecolor{whiteframe}{HTML}{ffffff}
\definecolor{pastelpink}{HTML}{ffb6c1}
\definecolor{creamframe}{HTML}{ffffcc}
\definecolor{lightlavender}{HTML}{e6e6fa}
\definecolor{pastelgray}{HTML}{cfcfc4}
\definecolor{pastelyellow}{HTML}{fdfd96}
\definecolor{lightgrayframe}{HTML}{d3d3d3}
\definecolor{lighterpastelblue}{HTML}{cde0e5}
\definecolor{lighterlavender}{HTML}{f3f3fd}  % Lighter lavender color
\definecolor{verylightpastelblue}{HTML}{e0f0f5}  % Even lighter pastel blue
\definecolor{verylightpastelblue}{HTML}{e0f0f5}  % Even lighter pastel blue
\definecolor{lightgrayframe}{HTML}{f0f0f0}       % Lighter gray
\definecolor{verylightpastelgreen}{HTML}{a3e4a7} % Even lighter pastel green
\definecolor{verylightwhiteframe}{HTML}{ffffff}  % Pure white
\definecolor{verylightpastelpink}{HTML}{ffdce0}  % Even lighter pastel pink
\definecolor{verylightcreamframe}{HTML}{ffffe6}  % Even lighter cream
\definecolor{verylightlavender}{HTML}{fbfaff}    % Even lighter lavender
\definecolor{verylightpastelgray}{HTML}{f0f0f0}  % Even lighter pastel gray
\definecolor{verylightpastelyellow}{HTML}{ffffc2} % Even lighter pastel yellow
\definecolor{verylightlightgrayframe}{HTML}{f5f5f5} % Even lighter light gray

\usepackage{listings}
\usepackage{xcolor}
\definecolor{burgundy}{RGB}{128,0,32}
\definecolor{lightburgundy}{RGB}{171,76,98}
\lstdefinestyle{SQLStyle}{
    language=SQL,
    basicstyle={\fontsize{6}{6}\ttfamily\selectfont},
    breaklines=true,
    breakatwhitespace=true,
    showstringspaces=false,
    keywordstyle=\color{blue},
    stringstyle=\color{lightburgundy},
    commentstyle=\color{green!60!black},
    numbers=none,
    tabsize=2,
    showtabs=false,
    frame=none,
    xleftmargin=0pt,
    aboveskip=5pt,
    belowskip=5pt
}
\definecolor{darkgreen}{RGB}{0, 100, 0}

\lstdefinestyle{VerilogStyle}{
    language=Verilog,
    basicstyle=\ttfamily\fontsize{6}{6}\selectfont, % Font size 6 for the entire code
    keywordstyle=\color{darkgreen}\bfseries, % Apply bold and dark green to customized keywords
    commentstyle=\color{gray},
    stringstyle=\color{red},
    numbers=left,
    numberstyle=\tiny,
    stepnumber=1,
    numbersep=5pt, % Space between numbers and code
    xleftmargin=10pt, % Left margin to align numbers within the tcolorbox
    framexleftmargin=10pt, % Padding for the line numbers inside the box
    breaklines=true,
    breakatwhitespace=false,
    showspaces=false,
    showstringspaces=false,
    showtabs=false,
    tabsize=2,
    morekeywords=[1]{module, endmodule, input, output, assign}, % Keywords to customize
}

\begin{document}

% title: hardaware design library querieing with natural language 
% use agent or asssistant 
% add LLM next to the agents names
% Dispatcher -> Dispatcher 
% interpolate the delay table 

\title{MetRex: A Benchmark for Verilog Code Metric Reasoning Using LLMs}

\author{Manar Abdelatty}
\email{manar_abdelatty@brown.edu}
% \author{Anonymus}
% \email{anonymus@anonymus.com}
\affiliation{%
  \institution{Brown University}
  \department{School of Engineering}
  \city{Providence}
  \state{RI}
  \country{USA}
}

\author{Jingxiao Ma}
\email{jingxiao_ma@brown.edu}
\affiliation{%
  \institution{Brown University}
  \department{School of Engineering}
  \city{Providence}
  \state{RI}
  \country{USA}
}

\author{Sherief Reda}
\email{sherief_reda@brown.edu}
\affiliation{%
  \institution{Brown University}
  \department{School of Engineering}
  \city{Providence}  \state{RI}
  \country{USA}
}

\begin{abstract}

Large Language Models (LLMs) have been applied to various hardware design tasks, including Verilog code generation, EDA tool scripting, and RTL bug fixing. Despite this extensive exploration, LLMs are yet to be used for the task of post-synthesis metric reasoning and estimation of HDL designs. In this paper, we assess the ability of LLMs to reason about post-synthesis metrics of Verilog designs. We introduce MetRex, a large-scale dataset comprising 25,868 Verilog HDL designs and their corresponding post-synthesis metrics, namely area, delay, and static power. MetRex incorporates a Chain of Thought (CoT) template to enhance LLMs' reasoning about these metrics. Extensive experiments show that Supervised Fine-Tuning (SFT) boosts the LLM's reasoning capabilities on average by 37.0\%, 25.3\%, and 25.7\% on the area, delay, and static power, respectively. While SFT improves performance on our benchmark, it remains far from achieving optimal results, especially on complex problems. Comparing to state-of-the-art regression models, our approach delivers accurate post-synthesis predictions for 17.4\% more designs (within a 5\% error margin), in addition to offering a 1.7x speedup by eliminating the need for pre-processing. This work lays the groundwork for advancing LLM-based Verilog code metric reasoning.

\end{abstract}



\maketitle

\section{Introduction}

Recent advancements in Large Language Models (LLMs) have demonstrated remarkable potential to transform the field of hardware design, across a wide range of tasks such as Verilog code generation~\citep{pearce2020dave, thakur2023benchmarking,Liu2023verilog,lu2024rtllm,thakur2023verigen}, EDA tools scripting~\citep{wu2024chateda}, designing AI accelerators~\citep{fu2023gpt4aigchip}, and fixing RTL syntax errors~\citep{tsai2023rtlfixer}. However, an area yet to be explored is the application of LLMs for reasoning and estimation of post-synthesis metrics of HDL designs. Given HDL code as input, LLMs could potentially infer gate-level details and estimate key metrics, such as area, delay, and static power.

% current datasets have no/less information about hw design; support by statistics ( how much verilog v.s python)
% DPO Tuning for LLMs, replacement for RL for human feedback
% mention that text is a lossless representation 

While current LLMs can generate raw Verilog code, they lack awareness of post-synthesis metrics and struggle to reason about them effectively. Prior works utilized LLMs to tweak Verilog code to meet area, delay, and power requirements using prompting methods~\citep{thorat2023advanced} or search methods like Monte-Carlo tree search~\citep{yao2024rtlrewriter}. However, these approaches mainly focus on refining Verilog code, and they do not fundamentally enhance the LLM's understanding of how different design choices impact post-synthesis metrics. Thus, there is a need for approaches that empower LLMs with deeper insights into the underlying relationships between HDL code and post-synthesis metrics.

In light of this, we introduce \emph{MetRex}, an LLM-based framework for high level metric estimation of HDL designs. \emph{MetRex} encompasses a large-scale dataset of $25,868$ HDL designs, each annotated with post-synthesis metrics on area, delay, and static power. To enhance the LLM's capability to understand and reason about these metrics, we propose a Chain of Thought (CoT) template that details the logical steps necessary for computing these metrics. To the best of our knowledge, \emph{MetRex} is the first framework that addresses the task of LLM-based code analysis for metric estimation of HDL designs. 

Our contributions are summarized as follows: 
\begin{itemize}[topsep=2pt, partopsep=1pt, leftmargin=*]
    \item We introduce a new dataset, \emph{MetRex}\footnote{https://github.com/scale-lab/MetRex}, for benchmarking Large Language Models (LLMs) for the task of reasoning about post-synthesis metrics of HDL designs. The dataset comprises 25,868 Verilog designs, each annotated with area, delay, and static power metrics. 

    \item We developed an automated flow using a Verilog compiler, a synthesis tool, and an LLM agent to detect and resolve syntax and synthesis errors, ensuring a dataset of clean, synthesizable designs.
    
    \item We introduce a Chain of Thought (CoT) prompting technique that improves the LLM's reasoning and understanding of post-synthesis metrics by 5.1\%, 5.4\%, and 8.9\% on the area, delay, and static power metrics, respectively, compared to direct prompting methods.

     \item We employ the \emph{MetRex} dataset in extensive Supervised Fine-Tuning (SFT) experiments, demonstrating that SFT can significantly improve the LLM performance in reasoning and estimating post-synthesis metrics on average by 37.0\%, 25.3\%, and 25.7\% on the area, delay, and static power, compared to few-shot prompting techniques.

    \item We compare the LLM estimation accuracy to regression-based models~\citep{wenj2023masterrtl}, highlighting their potential for this task in offering insightful and direct analysis of HDL code without the need for intermediary formats. LLMs improve the rate of obtaining accurate estimates within a 5\% error margin by 17.4\% while offering 1.7x faster analysis by eliminating the need for feature extraction and pre-processing. 
    
\end{itemize}

This paper is organized as follows. Section~\ref{related_work} discusses related work. Section~\ref{problem_formulation} presents a general problem formulation of the metric reasoning task with LLMs. Section~\ref{dataset} discusses the \emph{MetRex} dataset. Section~\ref{experiments} presents experimental results. Section~\ref{limitations} discusses current limitations and future directions. Finally, Section~\ref{conclusion} concludes the paper.

\section{Related Work}
\label{related_work}
Large Language Models (LLMs) have demonstrated strong potential in code generation tasks, where they can generate logically consistent code across various programming languages \citep{chen2021evaluating}. Their utility extends beyond mere code generation to include code reasoning and understanding, where they can repair bugs in codebases~\citep{jin2023inferfix}, reason about code execution~\citep{gu2024cruxeval}, and perform compiler optimizations~\citep{cummins2024meta}.

In the hardware domain, LLMs have been extensively applied to Verilog code generation~\citep{pearce2020dave, thakur2023benchmarking,Liu2023verilog,lu2024rtllm,thakur2023verigen}. VeriGen~\citep{thakur2023benchmarking, thakur2023verigen}, for instance, finetuned code LLMs for generating Verilog. Several benchmarks, such as RTLLM~\citep{lu2024rtllm} and VerilogEval~\citep{Liu2023verilog}, have been presented to standardize the evaluation of LLMs in Verilog code generation tasks. LLMs have also shown promise in code reasoning tasks, such as identifying and rectifying bugs in RTL designs~\citep{tsai2023rtlfixer} and optimizing Verilog code ~\citep{yao2024rtlrewriter}. Despite these advancements, the application of LLMs in reasoning about post-synthesis metrics of HDL designs remains largely unexplored. This represents a significant research opportunity, especially in enhancing the LLM's understanding of how different design choices impact post-synthesis metrics. 

The metric estimation of RTL designs has also been a subject of research for conventional machine learning techniques. Studies presented in~\citep{wenj2023masterrtl,Shao2014aladdin,zhang2023panda,Ouyang2023asap,Prianka2023earl} aim to provide an early estimate of the RTL post-synthesis or post-layout metrics to help accelerate the hardware design exploration process. These techniques typically transform the RTL design into different representational formats. MasterRTL~\citep{wenj2023masterrtl} proposes using simple operator graphs (SOGs) since it is closer to the synthesized netlist than abstract syntax trees (ASTs) \citep{Shao2014aladdin, zhang2023panda}. Manually engineered features are then extracted from these representations and used as inputs to regression-based models, such as XGBoost and graph neural networks, to predict post-synthesis metrics. 

In contrast, leveraging LLMs for this task offers a unique advantage. Unlike traditional methods, LLMs can process Verilog code directly, a lossless representation, thereby bypassing the need for manual feature extraction or transformation into intermediary formats. This direct processing enables LLMs to autonomously identify and extract features and patterns closely associated with synthesis outcomes, leading to potentially more insightful and faster analyses. Our study specifically aims to assess LLM reasoning capabilities about post-synthesis metrics of Verilog code, to broaden our understanding of the utility of these models in HDL design methodologies.

\section{Problem Formulation}
\label{problem_formulation}

In this section, we provide a general formulation of the HDL metric estimation task based on natural language instructions. Given a Verilog HDL design $\mathcal{V}$, the objective is to design an LLM-based model $f_{reason}$ to estimate the post-synthesis metrics $\mathcal{M}_{synth}$, where $\mathcal{M}_{synth}=f_{reason}(\mathcal{V})$. However, directly predicting the final metrics from HDL code is a complex reasoning task. This is mainly because LLMs are optimized for understanding and generating text in context rather than performing numerical calculations or predictions. LLMs are good at reasoning about a problem when it is described in words, but may struggle with abstract numerical predictions without explicit reasoning steps~\citep{wei2022cot}.

Therefore, we task the LLM with reasoning about the post-synthesis metrics through intermediate steps, $\mathcal{I}$, which include gate-level details from the synthesized netlist and natural language descriptions of how to calculate these metrics. These intermediate steps guide the LLM in a Chain of Thought (CoT) manner, helping it predict the final metrics. The CoT template includes information such as gate counts, area and power characteristics per gate, and critical path stages, all described in natural language. Thus, the task becomes predicting both the reasoning steps and the final post-synthesis metrics, expressed as $\langle(\mathcal{I}, \mathcal{M}_{\text{synth}})\rangle = f_{reason}(\mathcal{V})$. 

\section{MetRex Benchmark}
\label{dataset}

\subsection{Data Collection and Cleaning}
The initial phase in the dataset creation involved collecting a diverse range of HDL designs from various sources, as detailed in Table~\ref{data-sources}. We mainly relied on publicly available dataset sources that are used for evaluating LLMs for the task of Verilog code generation. Key sources include the RTL-Coder dataset~\citep{liu2023rtlcoder}, which contains designs generated by a GPT model, and VeriGen dataset~\citep{thakur2023verigen, thakur2023benchmarking}, which contains Verilog code extracted from GitHub repositories and academic textbooks. Additionally, we incorporated designs from ISCAS~\citep{Brglez1989iscas}, OpenCores~\citep{Albrecht2005iwls}, and NVLDA~\citep{nvdla_hw}. Altogether, we collected $25,868$ designs, the majority of which are self-contained modules. These designs form the training split of our dataset. For the test set, we derived it from the VerilogEval benchmark~\citep{Liu2023verilog}.
~\begin{table}[!t]
\fontsize{8}{8}\selectfont
\centering
% \resizebox{\columnwidth}{!}{%
\begin{threeparttable}
 \captionsetup{skip=0.5pt} 
 \caption{HDL design sources.}
 \label{data-sources}
 \begin{tabular}{lcc} 
 \toprule
Source  & Designs\tnote{1} & Complexity (Code Length) \\
         & (Count) & \{Min, Median, Max\} \\ 
\midrule
RTL-Coder\tnote{2}~\citep{liu2023rtlcoder} &  18,450 &  \{3, 29, 918\} \\
VeriGen~\citep{thakur2023verigen}  & 7,292 & \{5, 69, 27,025\}  \\
ISCAS'89~\citep{Brglez1989iscas} & 29  &  \{53, 530, 54,778\} \\
ISCAS'85~\citep{Brglez1989iscas} & 10  &  \{17, 1225, 3925\} \\
OpenCores~\citep{Albrecht2005iwls}  & 54  &  \{1, 103, 2716\} \\
NVLDA~\citep{nvdla_hw} & 33 &  \{19, 1333, 42,051\} \\
\midrule 
\multirow{2}{*}{\textbf{MetRex (ours)}} & 25,868 (Train) &  \multirow{2}{*}{\{3, 35, 54,778\}} \\
& 138\tnote{*} (Test) &  \\
 \bottomrule
 \end{tabular}
 \begin{tablenotes}
  \scriptsize
  \item[1] Number of designs after cleaning
\item[2] RTL-Coder dataset is LLM-generated. 
  \item[*] Test set derived from VerilogEval benchmark~\citep{Liu2023verilog}.
 \end{tablenotes}
\end{threeparttable}%
% }
\end{table}
\begin{figure}[!t]
      \captionsetup{skip=0.5pt} 
    \centering
     \begin{subfigure}{0.49\linewidth}
      \captionsetup{skip=1pt} 
        \centering
\includegraphics[width=\linewidth]{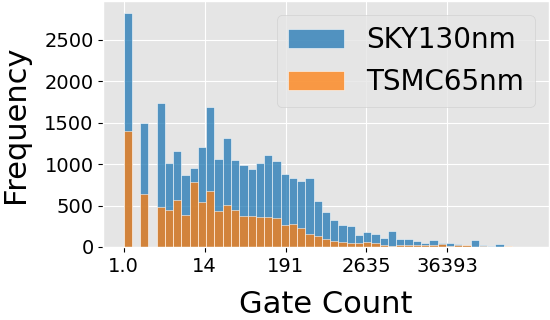} 
        \centering
        \caption{}
        \label{fig:gate_count}
    \end{subfigure}
    \begin{subfigure}{0.49\linewidth}
      \captionsetup{skip=1pt} 
        \centering
    \includegraphics[width=\linewidth]{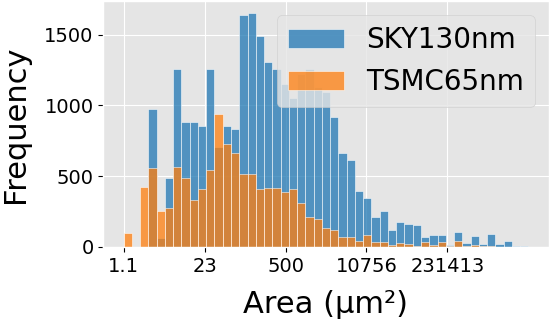}
        \centering
        \caption{}
        \label{fig:area}
    \end{subfigure}
  \begin{subfigure}{0.49\linewidth}
      \captionsetup{skip=1pt} 
        \centering
    \includegraphics[width=\linewidth, trim=0 0 10 0, clip]{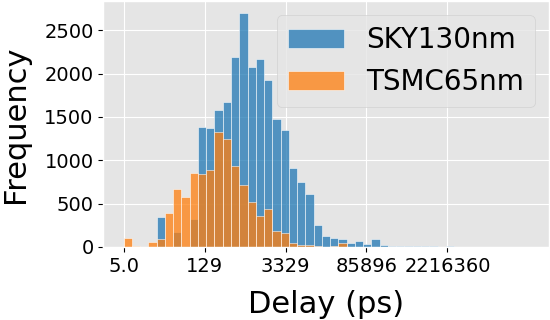}
        \centering
        \caption{}
        \label{fig:delay}
    \end{subfigure}
    \begin{subfigure}{0.49\linewidth}
      \captionsetup{skip=1pt} 
        \centering
        \includegraphics[width=\linewidth]{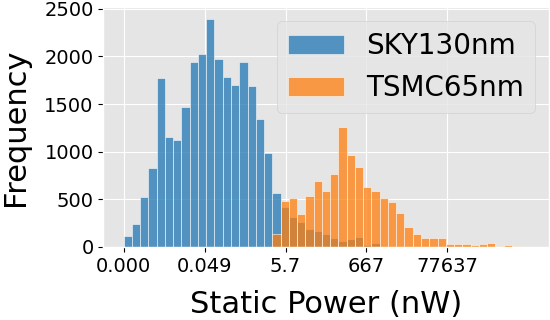}
        \centering
        \caption{}
        \label{fig:static_power}
    \end{subfigure}
    %  \begin{subfigure}{0.49\linewidth}
    %   \captionsetup{skip=1pt} 
    %     \centering
    % \includegraphics[width=\linewidth]{distribution/tokens1.png}
    %     \centering
    %     \caption{}
    %     \label{fig:num_tokens_1}
    % \end{subfigure}
    % \begin{subfigure}{0.49\linewidth}
    %   \captionsetup{skip=1pt} 
    %     \centering
    %     \includegraphics[width=\linewidth]{distribution/tokens2.png}
    %     \centering
    %     \caption{}
    %     \label{fig:num_tokens_2}
    % \end{subfigure}
\caption{Dataset analysis showing: (a) gate count, (b) area, (c) delay, (d) static power distribution for both Skywater 130nm and TSMC 65nm}
    \label{fig:data_metrics}
\end{figure}

Since the collected dataset contained LLM-generated and web-scraped Verilog code, it was important to clean the dataset to ensure the usability of these designs for further analysis and benchmarking. We undertook a comprehensive cleaning process that involved removing duplicate entries, filtering out non-synthesizable elements such as test benches and gate-level netlists, and rectifying errors and warnings detected during synthesis.

We automated the data cleaning flow by integrating an LLM agent with a synthesis tool and a Verilog compiler in an interactive feedback loop, inspired by the RTLFixer flow~\citep{tsai2023rtlfixer}. The LLM agent resolves errors and warnings flagged during synthesis or syntax checking. Warnings, such as unused signals, are included in the loop to prevent them from obscuring the relationship between Verilog code and post-synthesis metrics. Our workflow uses Icarus Verilog~\citep{williams2002iverilog}, Yosys~\citep{Wolf2013YosysAFV}, and Cadence Genus for syntax and synthesis verification.

Lastly, the cleaned dataset was taken through the synthesis flow to report the area, delay, and power metrics. We used Yosys~\citep{Wolf2013YosysAFV} for synthesizing the designs and reporting the area, and OpenSTA~\citep{Tutu2019openroad} for reporting the delay and power metrics. The designs were synthesized using the Skywater 130nm Process Design Kit (PDK)~\citep{skywater}. We also synthesized the designs using Cadence Genus Synthesis Solution and TSMC 65nm technology. Fig.~\ref{fig:gate_count} shows the gate count distribution after synthesis and Fig.~\ref{fig:data_metrics}b-d shows the distribution of the collected metrics in both technologies. 

~\begin{figure}[!t]
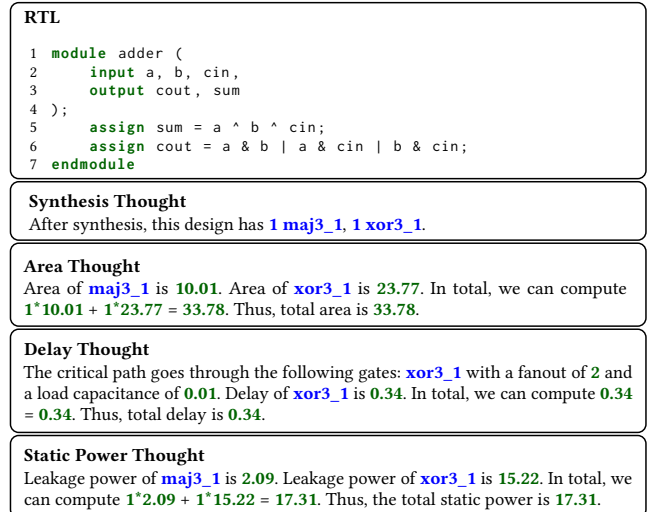

    \centering
    \captionsetup{skip=1.4pt} 
    % First Subfigure
    \begin{subfigure}{\linewidth}
        \centering
\begin{tcolorbox}[
    colback=white,
    colframe=black,
    fonttitle=\bfseries,
    width=\linewidth,
    boxrule=0.2mm, % Thinner boundary
    arc=1mm,
    fontupper={\fontsize{7}{8}\selectfont},
    left=2pt,
    right=4pt,
    top=0pt,
    bottom=-5.5pt,
    before skip=0pt,
    after skip=0.1pt
]
    \setlength{\parindent}{0pt}
    \setlength{\parskip}{0pt}
    \textbf{RTL}\\[-5pt]
    \fontsize{7}{8}\selectfont
    \begin{lstlisting}[style=VerilogStyle]
module adder (
    input a, b, cin,
    output cout, sum
);
    assign sum = a ^ b ^ cin;
    assign cout = a & b | a & cin | b & cin;
endmodule
    \end{lstlisting}
\end{tcolorbox}
\begin{tcolorbox}[
    colback=white,
    colframe=black,
    title=,
    fonttitle=\bfseries,
    width=\linewidth,
    boxrule=0.2mm, % Thinner boundary
    arc=1mm,
    fontupper={\fontsize{7}{8}\selectfont},
    left=4pt,
    right=1pt,
    top=1pt,
    bottom=-1pt,
    before skip=0.6pt,
    after skip=0.1pt
]
    \begingroup
    \setlength{\parindent}{0pt}
    \setlength{\parskip}{0.6pt}
    \textbf{Synthesis Thought}
    \par
    After synthesis, this design has 
    \textcolor{blue}{\textbf{1}} \textcolor{blue}{\textbf{maj3\_1}}, 
    \textcolor{blue}{\textbf{1}} \textcolor{blue}{\textbf{xor3\_1}}.
    \par
    \endgroup
\end{tcolorbox}
    \begin{tcolorbox}[
        colback=white,
        colframe=black,
        title=,
        fonttitle=\bfseries,
        width=\linewidth,
        boxrule=0.2mm, % Thinner boundary
        arc=1mm,
        fontupper={\fontsize{7}{8}\selectfont},
        left=2pt,
        right=4pt,
        top=2pt,
        bottom=-0pt,
        before skip=0.6pt,
        after skip=0.1pt
    ]
        \begingroup
        \setlength{\parindent}{0pt}
    \setlength{\parskip}{0.6pt}
        \textbf{Area Thought}
\par
Area of \textcolor{blue}{\textbf{maj3\_1}} is \textcolor{darkgreen}{\textbf{10.01}}. Area of \textcolor{blue}{\textbf{xor3\_1}} is \textcolor{darkgreen}{\textbf{23.77}}.
In total, we can compute \textcolor{darkgreen}{\textbf{1*10.01}} + \textcolor{darkgreen}{\textbf{1*23.77}} = \textcolor{darkgreen}{\textbf{33.78}}.
Thus, total area is \textcolor{darkgreen}{\textbf{33.78}}.
\par
        \endgroup
    \end{tcolorbox}
    
    % Delay Thought Box
    \begin{tcolorbox}[
        colback=white,
        colframe=black,
        title=,
        fonttitle=\bfseries,
        width=\linewidth,
        boxrule=0.2mm, % Thinner boundary
        arc=1mm,
        fontupper={\fontsize{7}{8}\selectfont},
        left=2pt,
        right=2pt,
        top=1pt,
        bottom=-0pt,
        before skip=0.6pt,
        after skip=0.1pt
    ]
        \begingroup
        \setlength{\parindent}{0pt}
    \setlength{\parskip}{0.6pt}
        \textbf{Delay Thought}
\par
The critical path goes through the following gates: \textcolor{blue}{\textbf{xor3\_1}} with a fanout of \textcolor{darkgreen}{\textbf{2}} and a load capacitance of \textcolor{darkgreen}{\textbf{0.01}}. Delay of \textcolor{blue}{\textbf{xor3\_1}} is \textcolor{darkgreen}{\textbf{0.34}}. 
In total, we can compute \textcolor{darkgreen}{\textbf{0.34}} = \textcolor{darkgreen}{\textbf{0.34}}.
Thus, total delay is \textcolor{darkgreen}{\textbf{0.34}}.
\par
        \endgroup
    \end{tcolorbox}
    
    % Static Power Thought Box
    \begin{tcolorbox}[
        colback=white,
        colframe=black,
        title=,
        fonttitle=\bfseries,
        width=\linewidth,
        boxrule=0.2mm, % Thinner boundary
        arc=1mm,
        fontupper={\fontsize{7}{8}\selectfont},
        left=2pt,
        right=2pt,
        top=1pt,
        bottom=-0pt,
        before skip=0.6pt,
        after skip=0.1pt
    ]
        \begingroup
        \setlength{\parindent}{0pt}
    \setlength{\parskip}{0.6pt}
        \textbf{Static Power Thought}
\par
Leakage power of \textcolor{blue}{\textbf{maj3\_1}} is \textcolor{darkgreen}{\textbf{2.09}}. Leakage power of \textcolor{blue}{\textbf{xor3\_1}} is \textcolor{darkgreen}{\textbf{15.22}}. 
In total, we can compute \textcolor{darkgreen}{\textbf{1*2.09}} + \textcolor{darkgreen}{\textbf{1*15.22}} = \textcolor{darkgreen}{\textbf{17.31}}. 
Thus, the total static power is \textcolor{darkgreen}{\textbf{17.31}}.
\par
        \endgroup
    \end{tcolorbox}
    \end{subfigure}    
\caption{Dataset sample, showing the Chain of Thought (CoT) template for estimating area, delay, and static power.}
\label{fig:sample}  
\end{figure}

\subsection{Chain of Thought (CoT) Template}

\begin{figure}[!t]
      \captionsetup{skip=0.6pt} 
    \centering
     \begin{subfigure}{0.49\linewidth}
      \captionsetup{skip=1pt} 
        \centering
    \includegraphics[width=\linewidth]{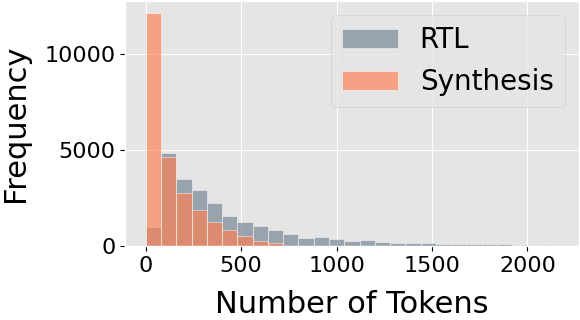}
        \centering
        \caption{}
        \label{fig:num_tokens_1}
    \end{subfigure}
    \begin{subfigure}{0.49\linewidth}
      \captionsetup{skip=1pt} 
        \centering
        \includegraphics[width=\linewidth]{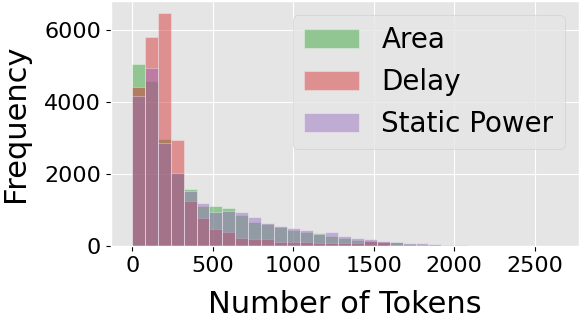}
        \centering
        \caption{}
        \label{fig:num_tokens_2}
    \end{subfigure}
\caption{Dataset analysis showing token count distribution for (a) RTL and synthesis thoughts and (b) area, delay, and static power thoughts in the Skywater 130nm dataset.}
    \label{fig:data_dist}
\end{figure}

To construct the intermediate reasoning steps for metric computation, we first parse the gates and their corresponding metrics from the synthesis reports. The parsed information is then used to construct natural language reasoning thoughts. Fig.~\ref{fig:sample} displays a sample from the \emph{MetRex} dataset of a full adder design synthesized using Skywater 130nm technology. The CoT template includes four primary reasoning thoughts: synthesis, area, delay, and static power. The synthesis thought includes a breakdown of the gate types and their count in the synthesized netlist. The area thought details the calculation of the total area by summing the individual areas of each gate type identified in the synthesis thought. The delay thought breaks down the stages of the critical path, including the type of gate, fanout, capacitive load, and delay for each stage. It then sums the delay per stage to calculate the total critical path delay. The static power thought outlines the leakage power for each gate type identified in the synthesis thought and sums these values to compute the total static power. Fig.~\ref{fig:num_tokens_1} shows the distribution of the number of tokens for the RTL and synthesis reasoning thought and Fig.~\ref{fig:num_tokens_2} shows the distribution of the number of tokens for the area, delay, and static power thoughts.

~\begin{table}[!t]
\centering
\captionsetup{skip=1pt} 
\fontsize{8}{8}\selectfont
\caption{Test dataset derived from the VerilogEval benchmark~\citep{Liu2023verilog}, categorized by difficulty level.}
\label{tab:test_data}
\begin{tabular}{>{\raggedleft}p{0.9cm} >{\raggedright\arraybackslash}p{3.4cm} >{\raggedright\arraybackslash}p{0.04cm} >{\centering\arraybackslash}p{2.5cm}}
\toprule
\textbf{Difficulty} & \textbf{Description} & \textbf{\# } & \textbf{Gate Count} \\
& & \textbf{} & \textbf{\{Min, Med, Max\}} \\
\midrule
\multirow{4}{*}{\shortstack{\textbf{Level-1} \\ \textbf{(L1)}}} 
  & Basic logic gates & 10 & \{1, 1, 1\} \\
  & Multi-bit gates & 9 & \{2, 2, 100\} \\
  & 1-bit comb. circuits & 5 & \{2, 2, 4\}\\
  & \textbf{Total \#} & \textbf{23} & \textbf{\{1, 2, 100\}} \\
\midrule
\multirow{6}{*}{\shortstack{\textbf{Level-2} \\ \textbf{(L2)}}} 
  & Adder circuits & 4 & \{2, 6, 15\}\\
  & Multi-bit comb. circuits & 23 & \{1, 3, 11\}\\
  & Flip-Flop registers  & 14 & \{1, 3, 24\} \\
  & Basic Seq. circuits  & 2 & \{8, 8, 8\}\\
  & \textbf{Total \#} & \textbf{43} & \textbf{\{1, 3, 24\}}\\
\midrule
\multirow{5}{*}{\shortstack{\textbf{Level-3} \\ \textbf{(L3)}}} 
  & Finite state machines & 24 & \{3, 11, 57\}\\
  & Counters  & 9 &  \{10, 14, 48\} \\
  & Complex comb. logic & 29 &  \{1, 7, 580\}\\
  & Advanced Seq. circuits & 9 &  \{11, 67, 607\}\\
  & \textbf{Total \#} & \textbf{72} &  \textbf{\{1, 14, 607\}} \\
\bottomrule
\end{tabular}
\end{table}

\section{Experimental Results}
\label{experiments}

\subsection{Evaluation Setup}
\label{eval_setup}

We conducted fine-tuning experiments using the train split of \emph{MetRex}, comprising $25,868$ designs and the Skywater version of the dataset for our experiments. Our evaluation set was derived from the VerilogEval benchmark~\citep{Liu2023verilog}, containing $138$ designs after excluding those with zero area or delay. The designs are categorized by difficulty in Table~\ref{tab:test_data}, where Level-1 includes simple combinational circuits with no more than 2-bit operators, Level-2 comprises moderate circuits such as adders, flip-flops, and shift registers, and Level-3 contains sophisticated designs like finite state machines and multi-bit arithmetic units. We measure performance using Mean Relative Error (MRE), defined in Eq.~\ref{eq:mre}, where $N$ is the number of designs in the test set, $\hat{R_{i}}$ is the LLM-estimated metric, and $R_{i}$ is the ground truth metric reported from the EDA tool.

% pass@k explanation: https://deepgram.com/learn/humaneval-llm-benchmark#the-passk-metric

% {\small
\begin{equation}
\label{eq:mre}
  MRE = \frac{1}{N}\Sigma_{i=1}^{N}\frac{|\hat{R_{i}}-R_{i}|}{R_{i}} \times 100\%
\end{equation}
% }

% However, the MRE has two main drawbacks: it can be heavily biased by outliers and it only evaluates the accuracy of the top-1 LLM-generated answer. Therefore, we introduce a new metric, \emph{acc@k}, inspired by the \emph{pass@k} metric~\citep{chen2021evaluating}, which is commonly used in evaluating LLMs for code generation. The \emph{acc@k} for an LLM measures the percentage of designs for which there is at least one generated solution among the top-$k$ samples that have an MRE within a specified margin $t$. The \emph{acc@k} is formally defined in Eq.~\ref{eq:pass@k}.

However, the MRE can be heavily biased by outliers and it only evaluates the accuracy of the top-1 LLM-generated answer. Therefore, we introduce a new metric, \emph{acc@k}, inspired by the \emph{pass@k} metric~\citep{chen2021evaluating}. The \emph{acc@k}, defined in Eq.~\ref{eq:pass@k}, measures the percentage of designs that meet a specific accuracy threshold when considering the median of the top-$k$ LLM predictions. Specifically, for each design in the test set, we compute the relative error between the ground truth (${R_i}$) and the median of the first $k$ predictions ($\hat{R}_{i,1:k}$), then check if this error falls within a specified margin $t$. 

% \begin{equation}
% \label{eq:pass@k}
%     acc@k(MRE \le t) = E_N\left[1-\frac{C(n-c, k)}{C(n,k)}\right]
% \end{equation}

{\small
\begin{equation}
\label{eq:pass@k}
acc@k(MRE \le t) = E_N\left[1\!\!1\left(\frac{|median(\hat{R}_{i,1:k}) - R_i|}{R_i} \le t\right)\right]
\end{equation}
}

% Here, $N$ is the total number of designs in the evaluation set, $n$ is the total number of samples generated by the LLM, and $c$ is the number of answers among the $n$ generated samples that fall within the MRE margin $t$, where $(n \gg k)$. We assess the \emph{acc@k} with $k$ values of 1, 5, 10, and MRE thresholds $t$ of $5\%$ and $10\%$. We set $n$, the number of generated solutions per problem, to $10$ because it is more practical in terms of response time and cost. A higher \emph{acc@k} signifies better model performance. 

Here, $N$ is the total number of designs in the evaluation set, and the indicator function  $1\!\!1(.)$  returns 1 if the relative error between the median prediction and the ground truth is within the margin $t$, and 0 otherwise. We assess the \emph{acc@k} with $k$ values of 1, 5, 10, and MRE thresholds $t$ of $10\%$ and $20\%$. A higher \emph{acc@k} indicates better model performance.

\begin{table}[!b]
\fontsize{8}{8}\selectfont
\centering
\captionsetup{skip=1pt} 
\caption{Impact of using chain-of-thought (CoT) prompt on the \emph{acc@5} of the test set, using in-context learning. $\times$ means without CoT, while $\checkmark$ means with CoT. The $\Delta$ values in \textbf{bold} represent the improvements from using the CoT prompt.}
\label{tab:cot_icl}
\begin{tabular}{>{\raggedright}p{0.8cm} >{\centering\arraybackslash}p{1.2cm} >{\centering\arraybackslash}p{1.2cm} >{\raggedleft\arraybackslash}p{0.8cm} >{\raggedleft\arraybackslash}p{0.8cm} >{\raggedleft\arraybackslash}p{1.6cm}}
\toprule
\multirow{2}{*}{\textbf{Margin}} & \multirow{2}{*}{\textbf{Model}} & \multirow{2}{*}{\textbf{With CoT?}} & \multicolumn{3}{c}{\textbf{acc@5 $\uparrow$}} \\
\cmidrule(lr){4-6}
 & & & \textbf{Area} & \textbf{Delay} & \textbf{Static Power} \\
\midrule
\multirow{6}{*}{\textbf{10\%}} & \multirow{3}{*}{\textbf{Mixtral-8x7b}} & $\times$ & 19.6\% & 19.6\% & 13.0\% \\
 & & $\checkmark$ & 19.6\% & 22.5\% & 18.8\% \\
 & & $\Delta$ & \textbf{+0.0\%} & \textbf{+2.9\%} & \textbf{+5.8\%} \\
\cmidrule(lr){2-6}
 & \multirow{3}{*}{\textbf{LLama3-8B}} & $\times$ & 10.1\% & 15.2\% & 6.5\% \\
 & & $\checkmark$ & 18.8\% & 23.9\% & 15.2\% \\
 & & $\Delta$ & \textbf{+8.7\%} & \textbf{+8.7\%} & \textbf{+8.7\%} \\
\midrule
\multirow{6}{*}{\textbf{20\%}} & \multirow{3}{*}{\textbf{Mixtral-8x7b}} & $\times$ & 25.4\% & 26.1\% & 15.9\% \\
 & & $\checkmark$ & 26.1\% & 29.0\% & 26.1\% \\
 & & $\Delta$ & \textbf{+0.7\%} & \textbf{+2.9\%} & \textbf{+10.1\%} \\
\cmidrule(lr){2-6}
 & \multirow{3}{*}{\textbf{LLama3-8B}} & $\times$ & 13.0\% & 21.0\% & 10.9\% \\
 & & $\checkmark$ & 23.9\% & 28.3\% & 21.7\% \\
 & & $\Delta$ & \textbf{+10.9\%} & \textbf{+7.2\%} & \textbf{+10.9\%} \\
\bottomrule
\end{tabular}
\end{table}

\begin{table*}[]
\centering
\fontsize{9}{9}\selectfont
\captionsetup{skip=1pt} 
\caption{Finetuning results using Skywater 130nm instruction datasets on the area, delay, and static power metrics. Results show improvements of supervised fine-tuning as measured by the \emph{acc@k} value, described in Section~\ref{eval_setup}. $\times$ means the model is not fine-tuned and uses only in-context learning, while $\checkmark$ means the model is fine-tuned using an instruction dataset of RTL code and metric reasoning pair. \textbf{Bolded} values highlight best-performing accuracy for a given metric and error margin.}
\setlength{\aboverulesep}{0pt}
\setlength{\belowrulesep}{0pt}
\setlength{\extrarowheight}{.5ex} 
\resizebox{\textwidth}{!}{%
\label{tab:sft}
\begin{tabular}{ccc|rrr|rrr|rrr}
\toprule
\multirow{2}{*}{\makecell{ \textbf{Margin (t)} \\ \textbf{MRE} $\le$ \textbf{t}}} & \multirow{2}{*}{\textbf{Model}} & \multirow{2}{*}{\textbf{Finetuned ?}} & \multicolumn{3}{c}{\textbf{Area (acc@k)} $\uparrow$} & \multicolumn{3}{c}{\textbf{Delay (acc@k) $\uparrow$}} & \multicolumn{3}{c}{\textbf{Static Power (acc@k) $\uparrow$}} \\ 
\cmidrule(lr){4-6} \cmidrule(lr){7-9} \cmidrule(lr){10-12}
 & & & \textbf{acc@1} & \textbf{acc@5} & \textbf{acc@10} & \textbf{acc@1} & \textbf{acc@5} & \textbf{acc@10} & \textbf{acc@1} & \textbf{acc@5} & \textbf{acc@10} \\
\midrule
\multirow{6}{*}{\textbf{10\%}} & \multirow{3}{*}{\textbf{Mixtral-MetRex-8x7b}} & $\times$ & 19.6\% & 19.6\% & 21.0\% & 23.9\% & 22.5\% & 22.5\% & 20.3\% & 18.8\% & 19.6\% \\
 &  & $\checkmark$ & 43.5\% & 46.4\% & 45.7\% & 42.8\% & 45.7\% & 45.7\% & 39.1\% & 39.9\% & 38.4\% \\
 &  & $\Delta$ & +23.9\% & +26.8\% & +24.6\% & +18.8\% & +23.2\% & +23.2\% & +18.8\% & +21.0\% & +18.8\% \\
\cmidrule(lr){2-12}
 & \multirow{3}{*}{\textbf{LLama3-MetRex-8b}} & $\times$ & 17.4\% & 18.8\% & 18.1\% & 20.3\% & 23.9\% & 22.5\% & 15.9\% & 15.2\% & 15.2\% \\
 &  & $\checkmark$ & \textbf{58.0\%} & \textbf{58.0\%} & \textbf{58.7\%} & \textbf{47.8\%} & \textbf{47.1\%} & \textbf{47.8\%} & \textbf{42.0\%} & \textbf{42.8\%} & \textbf{41.3\%} \\
 &  & $\Delta$ & +40.6\% & +39.1\% & +40.6\% & +27.5\% & +23.2\% & +25.4\% & +26.1\% & +27.5\% & +26.1\% \\
\midrule
\multirow{6}{*}{\textbf{20\%}} & \multirow{3}{*}{\textbf{Mixtral-MetRex-8x7b}} & $\times$ & 25.4\% & 26.1\% & 26.1\% & 31.9\% & 29.0\% & 29.7\% & 25.4\% & 26.1\% & 28.3\% \\
 &  & $\checkmark$ & 58.0\% & 61.6\% & 60.9\% & 50.7\% & 53.6\% & 55.8\% & \textbf{53.6\%} & \textbf{54.3\%} & \textbf{52.2\%} \\
 &  & $\Delta$ & +32.6\% & +35.5\% & +34.8\% & +18.8\% & +24.6\% & +26.1\% & +28.3\% & +28.3\% & +23.9\% \\
\cmidrule(lr){2-12}
 & \multirow{3}{*}{\textbf{LLama3-MetRex-8b}} & $\times$ & 22.5\% & 23.9\% & 22.5\% & 25.4\% & 28.3\% & 28.3\% & 22.5\% & 21.7\% & 21.0\% \\
 &  & $\checkmark$ & \textbf{73.2\%} & \textbf{76.1\%} & \textbf{74.6\%} & \textbf{61.6\%} & \textbf{64.5\%} & \textbf{63.8\%} & 52.2\% & 49.3\% & 47.1\% \\
 &  & $\Delta$ & +50.7\% & +52.2\% & +52.2\% & +36.2\% & +36.2\% & +35.5\% & +29.7\% & +27.5\% & +26.1\% \\
\bottomrule
\end{tabular}
}
\end{table*}

\subsection{In-Context Learning (ICL)}
\label{icl}

% talk about the LLMs used in ICL
% also mention that other name for ICL is few shot learning
% mention sampling temperature, gpt is using apis, mixtral and llama are quantized 4bits to be able to fit them in gpu memory, mre@1 for k=1. 

In-Context Learning (ICL), also known as few-shot prompting, is a prompting technique used to extrapolate LLM's knowledge to new tasks by learning from a small number of context-specific examples~\citep{dong2022survey}. In this study, we use ICL as a baseline to evaluate the base LLMs' ability to reason about post-synthesis metrics and to evaluate the impact of using chain-of-thought prompts on their reasoning capabilities. We use $10$ few-shot examples, mainly composed of RTL descriptions of basic gates such as AND, NOR, and OR, and their respective post-synthesis metrics. These examples are designed to enhance the LLM's understanding of how basic Verilog logic operators are translated into logic gates within the Standard Cell Library (SCL), including their area, delay, and power characteristics. 
% ~\input{tables/main_results2}

Using these 10 few-shot examples, we evaluated the ability of different LLMs to reason about the post-synthesis metrics of the test set. We ran the experiments in two modes: one using the chain-of-thought (CoT) template shown in Fig.~\ref{fig:sample}, and the second without CoT, where the total area, delay, static power of the few-shot examples are given directly without intermediate reasoning steps. The models tested include \texttt{Mixtral-8x7b}~\citep{mixtral2024mixtral97b}, and \texttt{Llama3-8b}~\citep{llama32024meta}. The models were run locally with 4-bit quantization on a single A6000 GPU and prompted at a sampling temperature of $0$. Results, summarized in Table~\ref{tab:cot_icl}, indicate that CoT prompting enhanced performance on average by 5.1\%, 5.4\%, and 8.9\% on the area, delay, and static power metrics respectively.

\subsection{Supervised Fine-tuning (SFT)}
While few-shot prompting can help extrapolate LLM knowledge to new tasks, it is limited by how many examples we can fit in the context window and does not intrinsically instill knowledge in the LLM weights. Supervised fine-tuning (SFT) can help align the LLM to a specific downstream task by adjusting the LLM weights according to an instruction dataset. Therefore, we conduct extensive experiments on supervised fine-tuning. We utilize the train split of the \emph{MetRex} dataset for instruction tuning and evaluate on the test dataset derived from the VerilogEval benchmark~\citep{Liu2023verilog}.

Our fine-tuning experiments are mainly focused on the \texttt{Mixtral-8x7b} and \texttt{Llama3-8b} models. We employ LoRA~\citep{hu2021lora}, a parameter-efficient fine-tuning technique that decomposes the weight matrices into smaller, manageable low-rank matrices. This approach significantly reduces the computational load and memory demands associated with fine-tuning LLMs. We fine-tune a LoRA adapter per metric using an instruction dataset of Verilog code and metric reasoning pair. Additionally, we quantize the LLMs to 4 bits to reduce the memory footprint. The models were fine-tuned using a maximum sequence length of $1048$ tokens on a single A40 GPU. 

Table~\ref{tab:sft} shows the \emph{acc@k} metric for the fine-tuned models using a LoRA rank of $128$. All LLMs are evaluated at a sampling temperature of $0.4$. Supervised fine-tuning significantly boosts the LLMs' estimation accuracy of the final metrics compared to their pre-trained few-shot prompted counterparts. SFT improved the \emph{acc@1} on average by 37.0\%, 25.3\%, and 25.7\% for the area, delay, and static power metrics, respectively. \texttt{Llama3-MetRex-8b} shows better performance on the area and delay metrics compared to \texttt{Mixtral-MetRex-8x7b}, which only outperforms \texttt{Llama3-MetRex-8b} in the static power estimates within the 20\% error margin.

\begin{figure*}[!t]
    \centering
    \captionsetup{skip=-1pt} 
    \begin{subfigure}[t]{\columnwidth}
        \centering
        \includegraphics[width=\columnwidth,trim=0 0 0 7, clip]{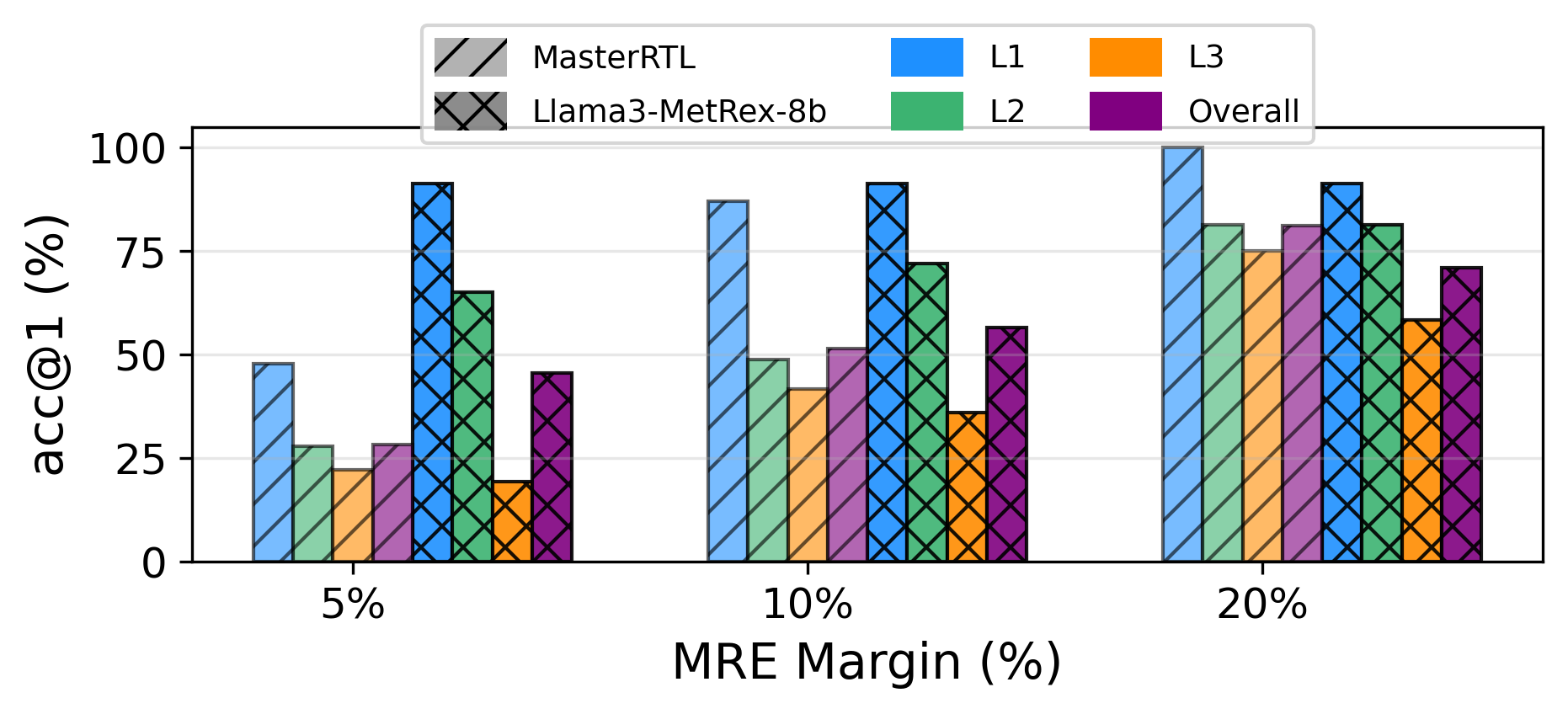}
    \captionsetup{skip=-1pt} 
        \caption{Area}
        \label{fig:area_accuracy}
    \end{subfigure}%
    \hfill
    \begin{subfigure}[t]{\columnwidth}
        \centering
        \includegraphics[width=\columnwidth, trim=0 0 0 7, clip]{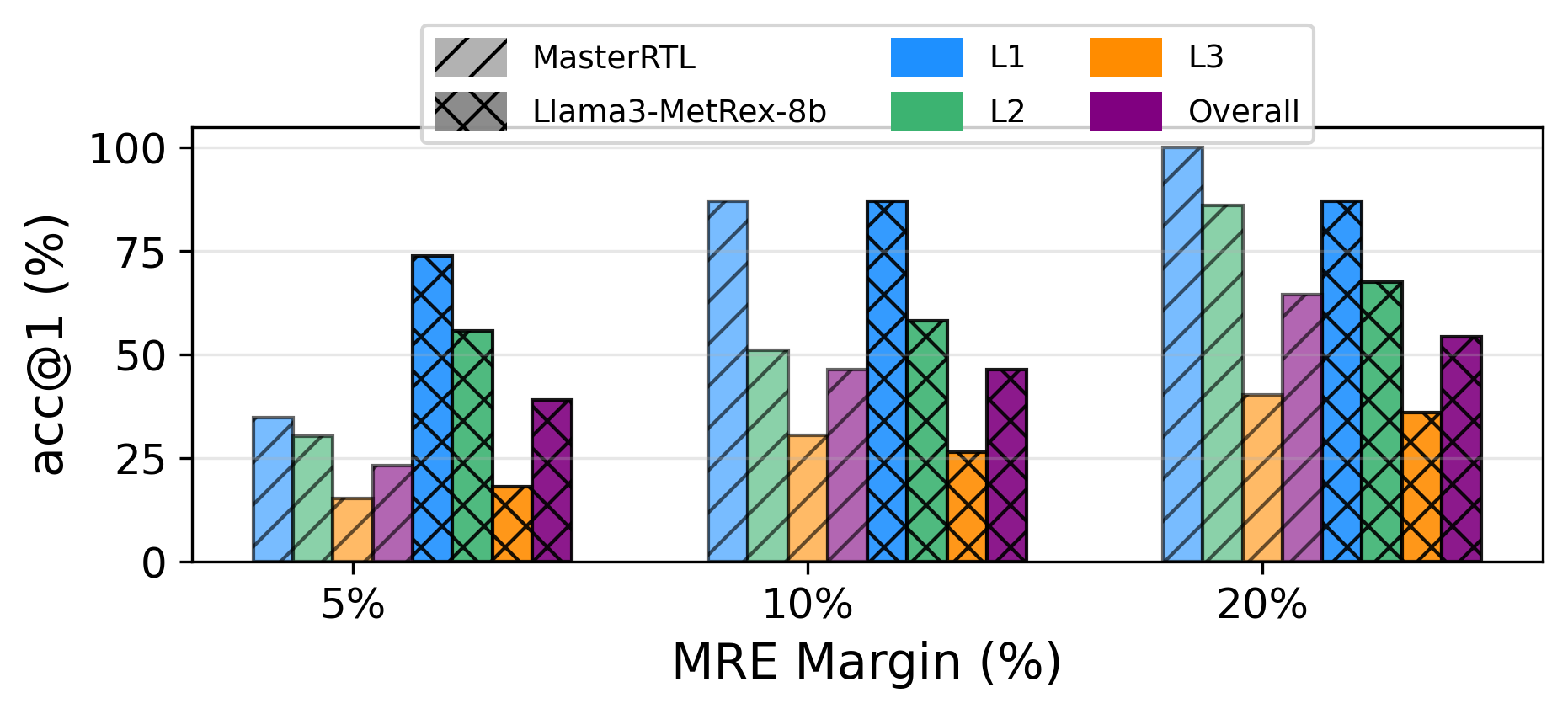}
    \captionsetup{skip=-1pt} 
        \caption{Static Power}
        \label{fig:static_power_accuracy}
    \end{subfigure}
    \captionsetup{skip=0pt} 
\caption{Comparison between MasterRTL~\citep{wenj2023masterrtl} and finetuned \texttt{Llama3-MetRex-8b}, showing (a) area and (b) static power accuracy across the three levels in the test set.}
    \label{fig:accuracy_comparison}
\end{figure*}

\subsection{Comparison to Regression-based Models}

In this section, we compare the accuracy of the finetuned \texttt{Llama3-MetRex-8b} (with a LoRA rank of $128$ and $256$ for area and static power, respectively) against regression-based machine learning approaches to highlight the opportunities and challenges of using LLMs for this task. We specifically compare to MasterRTL~\citep{wenj2023masterrtl}, which first converts the HDL code to a simple-operator graph (SOG) using Yosys, from which it extracts feature vectors for regression analysis. 

Fig.~\ref{fig:accuracy_comparison} shows the comparative performance for the area and static power estimation, highlighting the percentage of designs with MRE less than 5\%, 10\%, and 20\% across the three levels of complexity within our evaluation set. The data illustrates that the \texttt{Llama3-MetRex-8b} model can frequently generate more accurate answers than the regression-based model in less complex designs (level-1 and level-2) under the 5\% and 10\% error margins. However, it underperforms in level-3 primarily due to the increased reasoning complexity.

MasterRTL performs better under more relaxed error margins (20\%), mainly because it utilizes detailed gate-type features from the extracted SOG, which helps stabilize the performance of the regression model across a broad range of problem complexities. However, the ~\texttt{Llama3-MetRex-8b} model operates directly on Verilog code, resulting in higher sensitivity to variations in code design and susceptibility to generate extreme outliers. 

\begin{figure}[!t]
    \centering
    \captionsetup{skip=1pt} 
\includegraphics[width=\columnwidth,trim=0 6 6 5, clip]{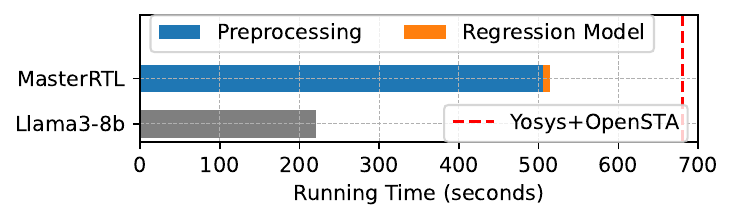}
    \caption{Run-time comparison between MasterRTL~\citep{wenj2023masterrtl} and \texttt{Llama3-MetRex-8b}. Evaluation is done using the H100 GPU. Pre-processing is done using an Intel Xeon CPU. }
\label{fig:runningtimes}
\end{figure}

Nonetheless, LLMs offer several advantages. First, they provide better overall estimates that are on average 17.4\% and 2.5\% more accurate within 5\% and 10\% error margins, respectively, compared to MasterRTL. Second, they provide interpretable results by explaining the breakdown of gates post-synthesis, offering insights beyond mere numerical predictions. Third, LLMs eliminate the need for preprocessing HDL code into intermediary formats and performing feature extraction, which significantly reduces runtime overhead. As shown in Fig.~\ref{fig:runningtimes}, the majority of MasterRTL's runtime is dedicated to generating the SOG and extracting feature vectors, totaling 505.3 seconds, whereas model inference time is minimal at just 8.5 seconds. Although LLMs inherently require substantial computational resources, leveraging GPU acceleration allows their runtime to be 2x faster than logic synthesis, and 1.7x faster than MasterRTL.

\section{Discussion and Future Work}
\label{limitations}
In this study, we aimed to assess the ability of LLMs to reason and predict post-synthesis metrics of Verilog code. Our collected dataset and evaluation framework mainly focused on self-contained and relatively small-scale designs, due to the limited fine-tuning context window. We aim to extend our dataset and fine-tuning experiments to include larger and more complex designs.

Additionally, we focused on the area, delay, and static power due to their relatively direct relationship to HDL code. Breaking down the switching power calculations in natural language to the LLM is more complicated as it requires propagating the activity factor through the logic gates and would require the LLM to understand the synthesized circuit topology and edge connection between these gates. Recent advancements in LLM research are showing progress in encoding graph data as natural language sequences~\citep{tang2023graphgpt,fatemi2023talk}, which will help with tackling the switching power reasoning. This could potentially improve the LLM accuracy on the delay estimation as well, as it will be able to reason about different paths in the circuit graph. Moreover, we assume that both the target technology node and synthesis strategy are fixed. We aim to investigate the influence of different synthesis strategies on the LLM estimation accuracy to offer insights into the utility of these models in different hardware design environments.  

Nonetheless, exploiting the reasoning capabilities of LLMs presents an exciting research opportunity for the hardware community. Particularly, the metric reasoning and estimation problem of HDL code will pave the way for tackling more difficult tasks such as generating efficient hardware code and accelerating the design exploration process. This study lays the groundwork for such future explorations. 

\section{Conclusion}
\label{conclusion}

In this paper, we introduced \emph{MetRex}, a new benchmark for evaluating LLMs for the task of reasoning about Verilog code post-synthesis metrics. \emph{MetRex} includes a large-scale dataset with a wide variety of HDL designs, annotated with their post-synthesis metrics, and chain of thought templates that detail the reasoning steps on how to compute these metrics. Our best performing model achieves accuracy rates of $73.2\%$, $61.6\%$, and $52.2\%$ when estimating area, delay, and static power metrics within a $20\%$ error margin, respectively. This work lays the foundation for a new line of research that leverages the capability of LLM reasoning to estimate post-synthesis metrics of HDL designs.

\section*{Acknowledgments}
This work is supported by NSF grant 2350180.

\newpage

% %%
% %% The acknowledgments section is defined using the "acks" environment
% %% (and NOT an unnumbered section). This ensures the proper
% %% identification of the section in the article metadata, and the
% %% consistent spelling of the heading.
% \begin{acks}
% \end{acks}

\bibliographystyle{unsrt}
\bibliography{acmart}

\end{document}